 \documentclass{vgtc}                 

\graphicspath{{figures/}{pictures/}{images/}{./}} 
\usepackage{amsmath}
\usepackage{times}                     
\usepackage{orcidlink}
\usepackage{tabu}                      
\usepackage{booktabs}  
\usepackage{cite}
\usepackage{lipsum}                    
\usepackage{mwe}                       
\usepackage{booktabs} 
\usepackage[]{todonotes}
\usepackage{tabularx}
\usepackage{comment}
\usepackage{enumitem}

\usepackage{mathptmx}                  

\usepackage{xcolor}
\definecolor{NavyBlue}{RGB}{0,0,128}

\onlineid{0000}

\vgtccategory{Research}

\title{Exploring Dimensions of Expertise in AR-Guided Psychomotor Tasks}

\author{Steven Yoo\thanks{e-mail: yoo.d@northeastern.edu}\\
        \scriptsize Northeastern University
\and Casper Harteveld\thanks{e-mail: c.harteveld@northeastern.edu}\\
     \scriptsize Northeastern University
\and Nicholas Wilson\thanks{e-mail: nicholas\_wilson@harvard.edu}\\
     \scriptsize Harvard University
\and Kemi Jona\thanks{e-mail: kjona@virginia.edu}\\
     \scriptsize University of Virginia
\and Mohsen Moghaddam\thanks{e-mail: m.moghaddam@northeastern.edu}\\
     \scriptsize \centering Northeastern University}

\teaser{
  \centering
  \includegraphics[width=\linewidth]{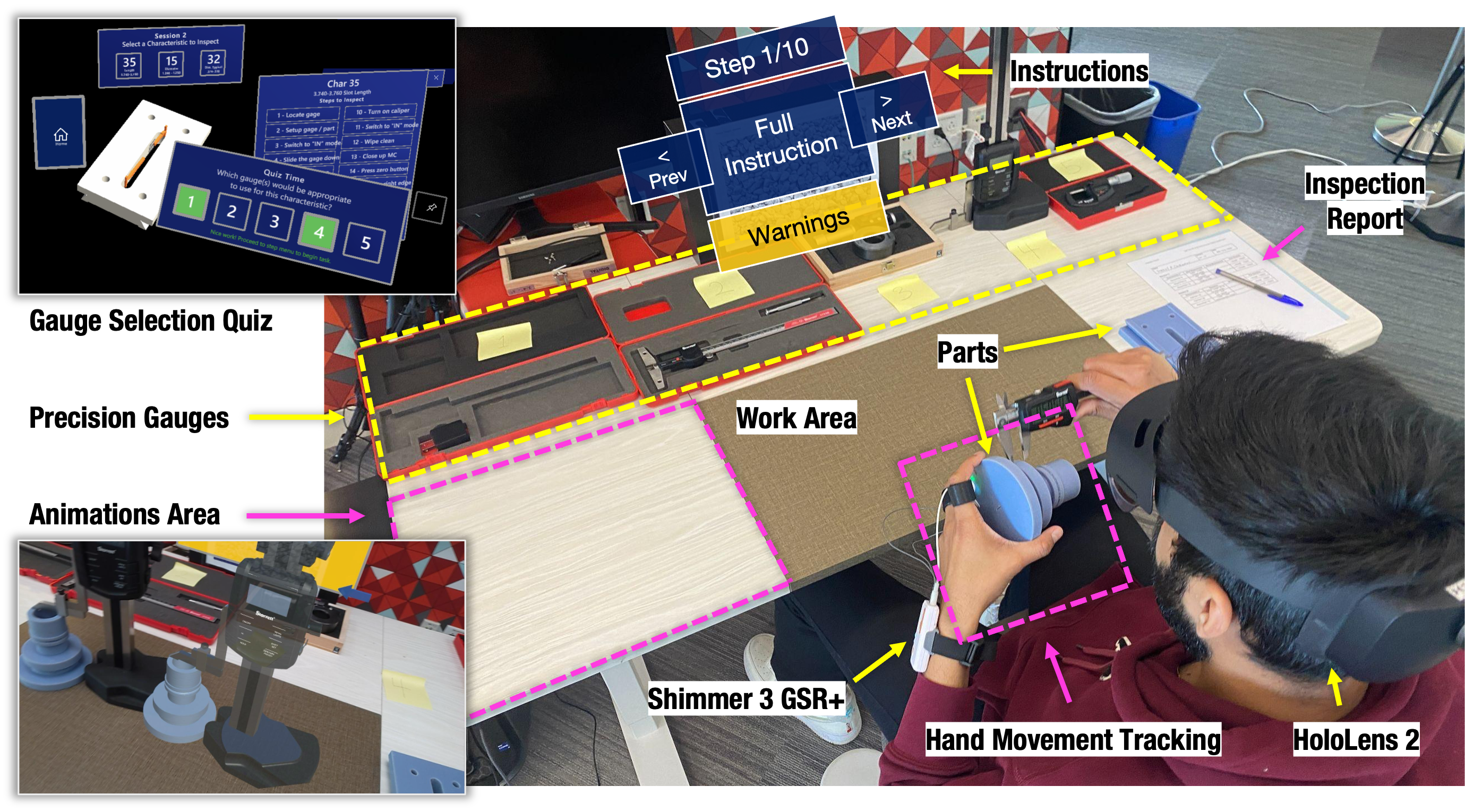}
  \caption{Exploring the decision making and proficiency of novices and experts in AR-guided tasks utilizing multimodal data.}
  \label{fig:teaser}
}

\abstract{
This study aimed to explore how novices and experts differ in performing complex psychomotor tasks guided by augmented reality (AR), focusing on decision-making and technical proficiency. Participants were divided into novice and expert groups based on a pre-questionnaire assessing their technical skills and theoretical knowledge of precision inspection. Participants completed a post-study questionnaire that evaluated cognitive load (NASA-TLX), self-efficacy, and experience with the HoloLens 2 and AR app, along with general feedback. We used multimodal data from AR devices and wearables, including hand tracking, galvanic skin response, and gaze tracking, to measure key performance metrics. We found that experts significantly outperformed novices in decision-making speed, efficiency, accuracy, and dexterity in the execution of technical tasks. Novices exhibited a positive correlation between perceived performance in the NASA-TLX and the GSR amplitude, indicating that higher perceived performance is associated with increased physiological stress responses. This study provides a foundation for designing multidimensional expertise estimation models to enable personalized industrial AR training systems.

} 


\begin{document}


\maketitle

\section{Introduction}
\label{intro}
Industrial tasks vary widely in complexity and skill requirements. Workers must master skills such as quick and accurate decision making, knowledge transfer between tasks, and precise task execution techniques \cite{eswaran2024augmented}. AR technology has proven to be effective in developing these skills in workers by providing contextual information in real time directly in their field of vision \cite{bottani2019augmented, lai2020smart}. However, most existing industrial AR systems often offer generic instructions that overlook the knowledge and individual needs of the worker \cite{lai2020smart}, thus limiting the personalization of the intervention. Given the abundance of rich multimodal data accessible through AR devices and wearables, it is feasible to integrate new inference mechanisms into industrial AR systems that dynamically estimate and adapt to the worker's expertise level during task performance. A key prerequisite is understanding different dimensions of expertise in these contexts, along with data-driven metrics that enable the estimation and adaptation of real-time expertise.

Previous studies have aimed to estimate expertise levels and distinguish between novice and expert behavior patterns across various tasks and dimensions of expertise. Examples include the dexterity of surgeons in operating rooms \cite{uemura2014analysis, louridas2016predictive, vajsbaher2022development}, troubleshooting by technicians in industrial settings \cite{farrington2006nature}, and rapid and accurate decision making by pilots \cite{schriver2017expertise, vidulich2010information}. However, the specific patterns and features of this expertise---and the extent to which they enhance task performance---are yet to be fully explored, especially in the context of AR-guided psychomotor tasks. There is a critical need to explore how the existing knowledge about the quantification and estimation of expertise in psychomotor tasks translates into AR technologies for industrial training and assistance \cite{moghaddam2021exploring}. This involves investigating how expertise-related factors influence interactions with AR interfaces, decision-making processes, and skill acquisition within AR-guided environments. By addressing this gap, we can unlock the full potential of AR in enhancing training effectiveness and performance in psychomotor tasks, which involves the aforementioned personalized and customized AR systems.

In this paper, we investigate multiple hypotheses under two key dimensions of expertise to further our understanding of the nuances of novice and expert performance of industrial psychomotor tasks. (1) \textit{Decision making dimension}: Experts significantly outperform novices in making fast and correct decisions, more independently and confidently, and are more capable of effectively transferring their knowledge to new tasks and scenarios. (2) \textit{Technical proficiency dimension}: Experts show higher degrees of proficiency in performing psychomotor tasks characterized by greater accuracy, dexterity, and efficiency compared to novices. We use the term ``dimension'' to emphasize that these are measurable indicators that can vary independently. For example, a novice surgeon may be the same as an expert in selecting the right instruments during an operation (decision making) but lack the same dexterity in using them (technical proficiency). We also acknowledge that this list is not comprehensive and that other dimensions may exist and should be considered in future studies.  

We tested the hypotheses by examining the performance of novices and experts in AR-guided psychomotor tasks, using multimodal data captured from AR devices and wearables, including hand tracking data, GSR measurements, and visual gaze information. We focus our experiments on precision inspection as an industrial psychomotor task, which involves (1) selecting measurement gauges based on the characteristics and tolerances of the parts, as well as the configuration and precision of the gauges, (2) preparing the selected gauge for inspection through initial setup, cleaning, and calibration, and (3)  measuring the characteristic of interest and accepting or rejecting the part. Our rationale for choosing this use case is the inherent complexity of the task and the need for critical thinking, decision making, and technical proficiency, which provides measurable performance metrics applicable to broader industrial tasks. The purpose of this paper is to lay the foundation for designing expertise estimation models informed by expertise dimensions in AR-guided psychomotor tasks and the data modalities that facilitate their measurement. This is a crucial step towards enabling industrial AR applications capable of dynamically adapting content and providing learning and accessibility interventions to the worker's level of expertise.

\section{Related Work}
Expertise is a multifaceted concept that involves a variety of skills, knowledge, and competencies acquired through training and experience in specific domains. An accurate assessment of an individual's level of expertise from task performance metrics and physiological signals can enable personalized training and assistance \cite{yoo2023augmenting}. This is crucial for complex psychomotor tasks that involve coordinated mental processes and physical movements, such as playing instruments, performing surgery, driving, or operating equipment \cite{hofstad2013study, willis2012comparing}. Recent advances in AR technology have opened new avenues for improving the learning of these and similar complex tasks, particularly in industry \cite{bottani2019augmented}. Further advances require a deeper understanding of the interplay between expertise and task performance.

Previous studies highlight the distinctions between novices and experts in different fields. Experts demonstrate superior performance marked by efficient information processing, pattern recognition, and adaptive decision-making abilities, which enable them to navigate complex situations with greater ease and precision \cite{ericsson2008expertise, feltovich2006studies}. In contrast, novices often rely on analytical reasoning and explicit rule-based knowledge, resulting in slower and less effective task execution \cite{klein1997naturalistic}. This contrast manifests itself in various psychomotor domains, ranging from surgery \cite{rosendal2023technical} to aviation \cite{bruder2019differences} and music performance \cite{brown2015expert}, underscoring the universal nature of the acquisition of expertise and its implications for the optimization of training and performance. Several studies have attempted to analyze and measure the differences between novices and experts from various perspectives, including:

\begin{itemize}
[itemsep=-0.25ex]
    \item \textit{Gaze behavior}: Evidence suggests that experts focus longer on relevant areas, indicating deeper cognitive processing and greater familiarity with the task and its environment, while novices exhibit more pronounced scattered gaze patterns and scanning behavior \cite{yoo2023modeling, li2023using}. In aviation, for example, experienced pilots exhibit shorter dwell times, more total fixations, fewer altimeter fixations, and better-defined eye-scanning patterns compared to novices \cite{li2013investigation,ziv2016gaze}. These findings concur with the information reduction hypothesis, which posits that experts optimize the amount of information processed by selectively allocating their attention to task-relevant stimuli and ignoring irrelevant stimuli \cite{castet2000motion, brams2019relationship}. 

    \item \textit{Adaptability}: Experts are shown to demonstrate better adaptability and resourcefulness, flexibly and creatively applying their knowledge in handling novel situations and solving new problems \cite{carbonell2014experts}. The ability to dynamically adjust strategies in response to unique challenges is crucial, not only for individual experts but also for teams in high-stakes settings. Evidence suggests the role of expertise in maintaining performance through strategic planning and role flexibility, especially in scenarios like emergency management where improvisation is necessary \cite{eaton2019linking,mendonca2007cognitive}. 

    \item \textit{Stress management}: Physiological and cognitive differences are reported between novices and experts during stress-induced decision making. Stress is shown to considerably impair the performance of novice surgeons \cite{arora2010stress}. Significant differences between novices and experts are also identified in their stress management, performance under pressure, use of mental practice, and concentration abilities \cite{arora2011mental}. In contrast to experts, who often maintain their performance under stress, novices show higher physiological stress markers, leading to less effective decision making as they rely more on learned rules and are prone to biases \cite{fooken2016role, harms2017stress}.

    \item \textit{Dexterity}: Several studies have focused on comparing novice and expert patterns of performance in terms of dexterity in fields such as surgery, industrial troubleshooting, and aviation \cite{louridas2016predictive, vajsbaher2022development, farrington2006nature, schriver2017expertise, vidulich2010information}. For instance, expert surgeons are shown to exhibit smoother and more stable hand motions compared to novices, and have more coherent and consistent movements, with greater long-range stability, while novices show more erratic and less controlled hand motions \cite{uemura2014analysis}. This indicates that the proficiency of expert surgeons is reflected in the refined stability and coherence of their hand movements.
\end{itemize}

Despite the considerable progress reported in this section on the differences between novices and experts, there remain gaps in understanding the interplay between expertise and task performance, and especially how this can be translated to AR applications. Understanding this interplay in contexts where AR applications will be used can lead to more effective and adaptive AR systems that tailor instructions and interventions to varying expertise levels. 

\section{Expertise Dimensions and Hypotheses}

We systematically explore the specific aspects of behavior and performance that characterize expert and novice interactions in complex psychomotor tasks. The study is structured around two main dimensions of expertise: decision-making and technical proficiency. The former involves the speed, accuracy, independence, and adaptability of a worker in making decisions. The latter focuses on accuracy, dexterity, and efficiency in executing the task. This section introduces these dimensions along with their respective metrics and hypotheses.

\subsection{Decision Making Dimension}
The first dimension examines whether there is a notable difference between experts and novices in their ability to make quick, accurate, and independent decisions. It also explores whether experts demonstrate superior capability in transferring their expertise to novel tasks and situations. For instance, a key decision in our precision inspection study is selecting the appropriate gauge based on the part's characteristics and required tolerances, as well as the gauge's configuration and precision.

\paragraph{Metrics:} \textit{Speed}: Time taken to reach a decision. \textit{Correctness}: Whether the decision made is correct or not. \textit{Independence}: Degree of reliance on external support for decision making. \textit{Adaptation}: Ability to apply skills to new and varied tasks.

\paragraph{Hypotheses:} \textit{H1}: The correctness of decisions significantly differentiates novice and expert performance in psychomotor tasks. \textit{H2}: There are significant differences in the time that experts and novices spend scanning their environment before making a decision. \textit{H3}: Novices and experts differ in their help-seeking behavior and reliance on detailed instructions during tasks. \textit{H4}: Previous experiences influence decision-making, tool selection, and adaptation to new tasks among novices and experts. \textit{H5}: Experts exhibit lower degrees of stress than novices in decision making.

\subsection{Technical Proficiency Dimension}
The second dimension explores whether experts demonstrate higher levels of proficiency in performing psychomotor tasks in terms of accuracy, dexterity, and efficiency compared to novices. The goal is to assess the technical proficiency of both experts and novices by measuring their performance using metrics related to precision, manual coordination, and efficiency in task completion. In the context of our precision inspection use case, technical proficiency involves the worker's ability to use the selected gauge efficiently and accurately to make accurate measurements.

\paragraph{Metrics:} \textit{Accuracy}: Quality of task execution and inspection outcomes. \textit{Dexterity}: Continuity of hand movements---the frequency and duration of hesitations/pauses during task performance. \textit{Efficiency}: The amount of time spent to complete the task.

\paragraph{Hypotheses:} \textit{H6}: There is a significant difference between task execution by novices and experts in terms of the accuracy of inspection report results. \textit{H7}: Hand movement patterns differ between novices and experts in the execution of psychomotor tasks. \textit{H8}: The efficiency of task execution, measured by the time taken to complete tasks, varies between experts and novices.

\section{User Study}
This section details the study materials and data collection procedures.

\subsection{Participants}
Twenty participants (4 females, 16 males; mechanical/industrial engineering and arts/science students) were recruited based on their experience with inspection gauge tasks. The participants were divided into novice and expert groups based on a pre-questionnaire that assessed their technical skills and theoretical knowledge of precision inspection. The study was approved by Northeastern University's Institutional Review Board. 

\paragraph{Novice Group:} This group included 10 students from non-engineering fields with limited experience in precision inspection using gauges. Familiarity ratings ranged from 1 (not familiar at all) to 5 (strong familiarity), averaging $M = 1.0$, $SD = 0.94$. The group consisted of 6 males and 4 females, including 5 master's students, 3 PhD students, and 2 undergraduates. Commonly used gauges were the Electronic Caliper, Depth Gauge, and Height Gauge, each used by 3 participants. 

\paragraph{Expert Group:} This group included 10 engineering students with strong experience in precision inspection using gauges. Familiarity ratings ranged from 1 to 5, averaging $M = 4.1$, $SD = 0.74$. The group consisted of 1 junior undergraduate and 9 master's students, with 5 participants having relevant industry experience. Commonly used gauges were the Electronic Caliper, Height Gauge, and Electronic Micrometer, each used by 9 participants.

\begin{figure}[t!]
\center
\includegraphics[width=\linewidth]{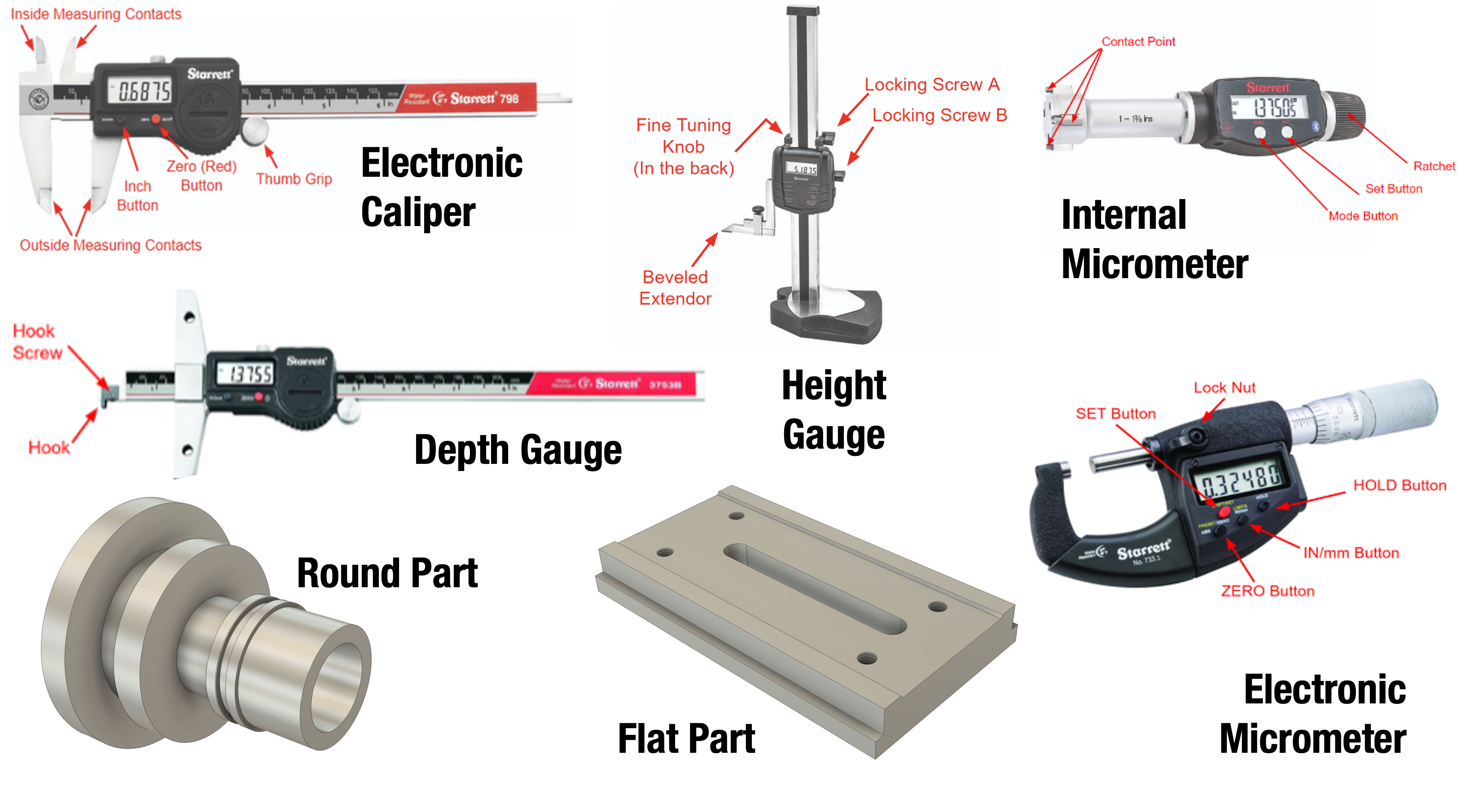}
\caption{Study's precision inspection gauges and parts.}
\label{gauges_parts}
\end{figure}

\subsection{Task and Apparatus}
We conducted a between-subjects study on the AR-guided inspection of 3D printed parts using various gauges (Figure~\ref{gauges_parts}). The task involved selecting appropriate gauges based on part geometries and required tolerances, as well as the gauge configurations and precisions. The participants were responsible for setting up and calibrating the gauges for inspection, locating characteristics on the parts, storing the gauge on the part, and measuring the respective characteristic. They reported measurement values and decided whether to accept or reject them based on the prescribed geometries and tolerances. The inspection accuracy was evaluated using binary scoring.

\begin{figure*}[htb]
\centering
\includegraphics[width=0.495\textwidth]{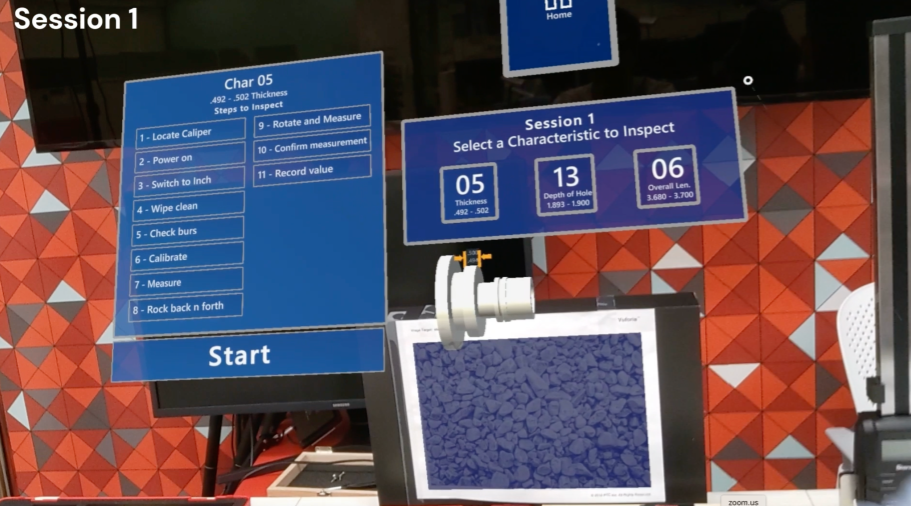}
\includegraphics[width=0.495\textwidth]{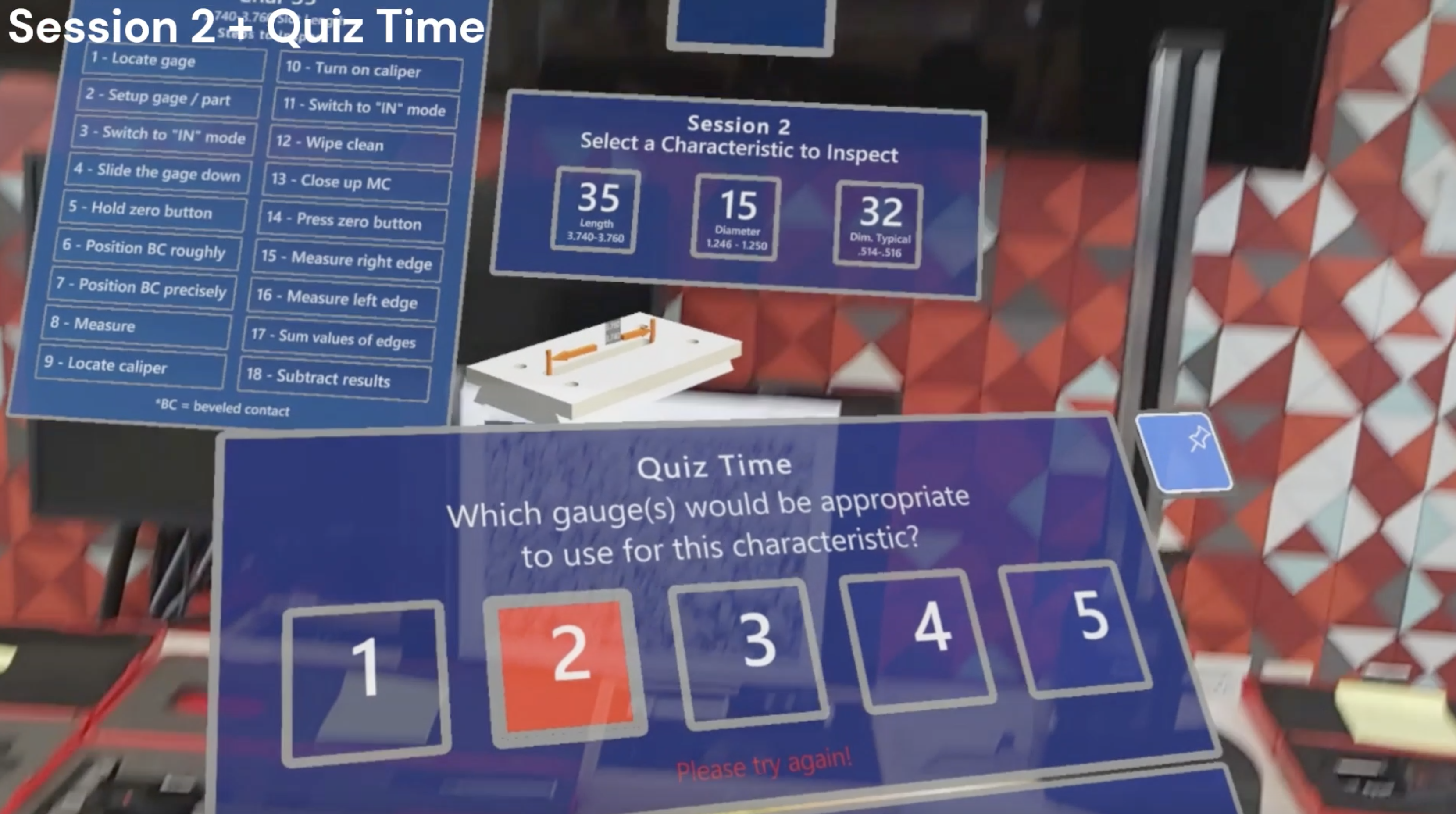}
\caption{AR application interface (HoloLens 2) for inspection tasks.
Left: Interface for Session 1, where participants follow detailed instructions and animations for Characteristics 5, 6, and 13 using specific gauges. This session emphasizes step-by-step guidance with full instructions always visible.
Right: Interface for Session 2, featuring the Gauge Selection Quiz. Participants choose appropriate gauges and measure Characteristics 15, 32, and 35 with optional access to full instructions. Each participant has three attempts per characteristic, with feedback provided after incorrect responses.}
\label{session1-2}
\end{figure*}

\paragraph{Session 1: Round Part with Instructions and Animation.}
In Session 1, participants were guided to use specific gauges for various characteristics with detailed instructions and animations. For example, the electronic caliper was used to measure Characteristic 5 (0.492-0.502 in, thickness), the depth gauge for Characteristic 13 (1.893-1.900 in, depth of hole) and the height gauge for Characteristic 6 (3.680-3.700 in, overall length). The app provided interactive buttons for each characteristic, which activated step-by-step instructions and animations to guide users through the inspection process (Figure~\ref{session1-2}). This session aimed to introduce users to simpler, more straightforward tasks using familiar and easier gauges. The process started with short instructions and a start button. Upon initiating the task, the system transitioned to full instructions accompanied by animations, ensuring that participants received full guidance throughout the session.


\paragraph{Session 2: Round and Flat Part with Quiz Module.}
In Session 2, participants selected appropriate gauges to measure characteristics based on physical configurations and tolerances. For example, Characteristic 35 (3.740-3.760 in, center-to-center slot length) was measured with both the height gauge and the electronic caliper, Characteristic 15 (1.246-1.250 in, inside diameter of hole) with the internal micrometer, and Characteristic 32 (0.514-0.516 in, thickness) with the electronic micrometer. This session focused on decision making, which required participants to measure both flat and round parts. Session 1, Session 2 presented short descriptions by default. The participants had the option to access the full instructions and animations within the app as needed (Figure~\ref{session1-2}). The design of this session reflected the increased complexity of the tasks, encouraging a more active engagement with the instructions. The participants also had three attempts per characteristic and received immediate feedback after incorrect responses, reinforcing the decision-making aspect of the session.


\subsection{Procedure}

The study was conducted sequentially, consisting of two sessions separated by a brief break. The procedure followed the following steps: Participants began by completing pre-questionnaires to collect baseline information, including demographics and prior experience with AR and HoloLens 2. Following the pre-questionnaires, participants received an introduction to the task and the AR application, outlining the objectives and the procedure of the study. The participants then participated in a 10-15 minute training session with the HoloLens Tips app to become familiar with the device. During the experiment, the Shimmer GSR+ was calibrated to ensure accurate physiological data collection. The first session commenced immediately after the training. During this session, participants interacted with the AR application using hand gestures, which allowed them to access AR content such as instructions, animations, and safety checklists. Gaze data, including gaze origin, direction, and hit information, were recorded using the HoloLens 2. After Session 1, participants took a 10-minute break to rest before the second session. Participants were also allowed to take a longer break if necessary. The second session started immediately after the break and followed the same procedure as Session 1, with continued interaction with the AR application and data collection. Upon completing both sessions, participants completed a post-study questionnaire. This included NASA-TLX to assess cognitive load, as well as questions regarding self-efficacy, user experience with the HoloLens and AR application, and general feedback on the study. NASA-TLX was administered once at the end of the entire study to assess the cognitive load experienced in both sessions. Data were collected throughout both sessions using the HoloLens 2 and Shimmer3 GSR+ at a 128 Hz sampling rate. Gaze tracking, captured at a 30 Hz sampling rate using the HoloLens Mixed Reality Toolkit (MRTK), included detailed data on gaze origin, direction, and hit points, providing insight into the specific targets (e.g., electronic calipers, height gauges, and quiz modules) participants focused on. The precision of the participants' decisions was evaluated using binary scoring.

\begin{figure*}[htb]
\centering
\includegraphics[width=0.42\textwidth]{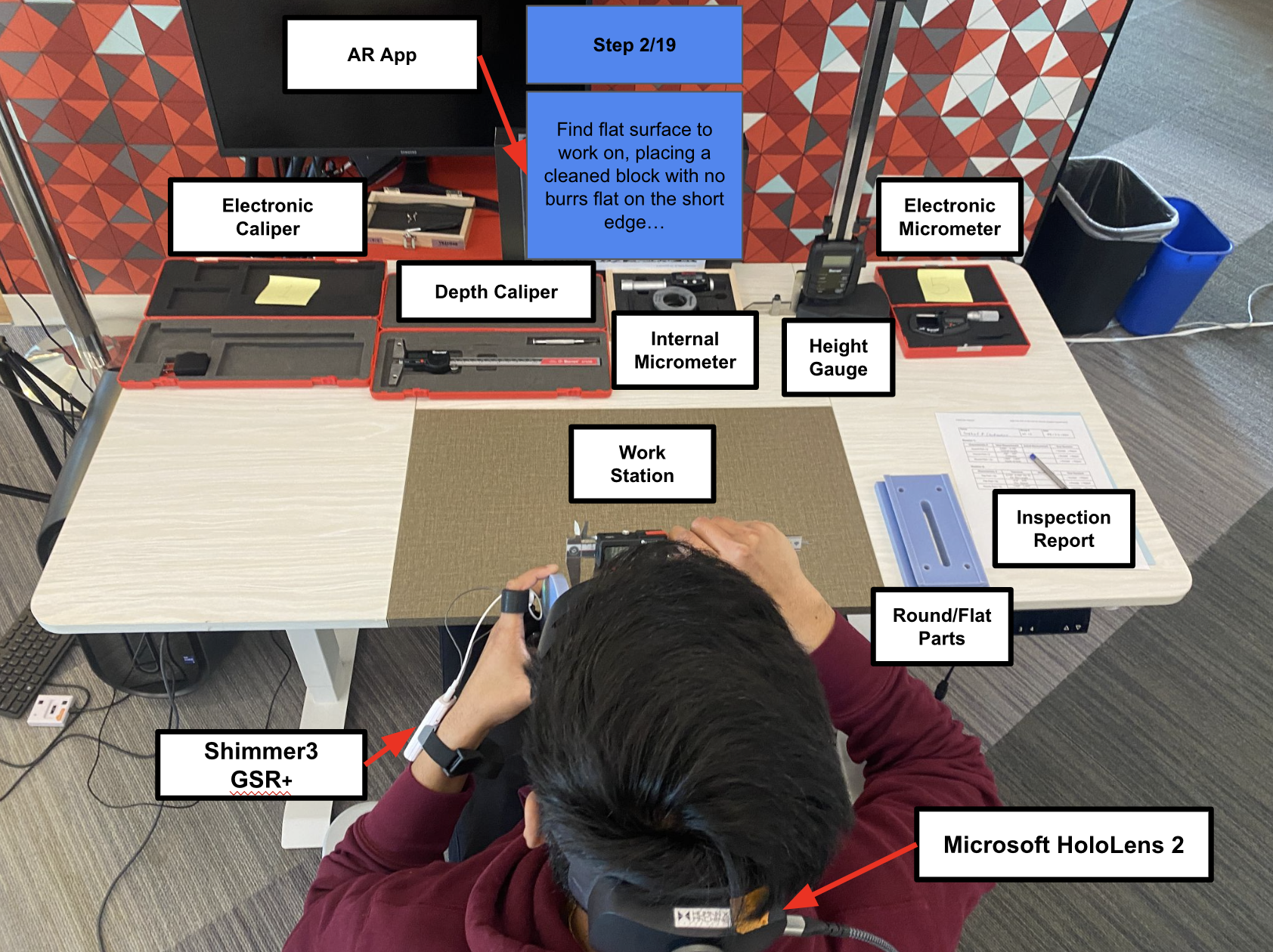}
\includegraphics[width=0.51\textwidth]{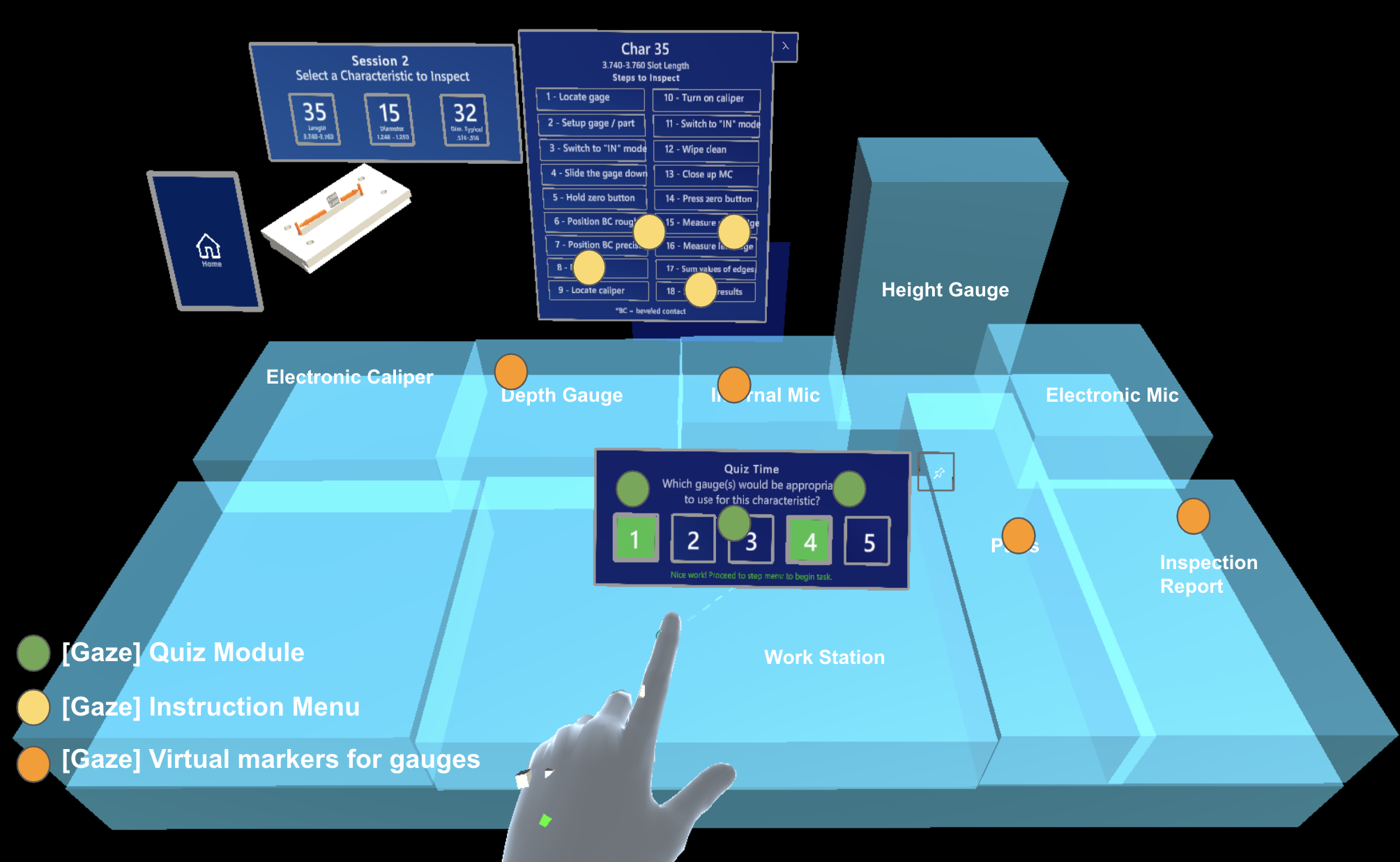}
\caption{Left: Real-world view and AR application with visual scanning areas. Right: Blue highlighted regions indicate visual scanning areas, tracking gaze on different components.}
\label{visual-gaze}
\end{figure*}

\section{Results}

This section presents the experimental results organized according to the eight hypotheses linked to the two dimensions of expertise.

\subsection{Correctness of Decisions (H1)}

We analyze the differences between the correctness of decisions made by experts and novices, as indicated by the number of attempts to complete tasks correctly. During Session 2, all participants completed a multiple choice quiz module before beginning their tasks, designed to assess their knowledge and ability to properly operate inspection gauges. The participants had up to three attempts to select the correct gauge; after three incorrect attempts, the AR system highlighted the correct gauge. Fewer attempts indicate greater proficiency and more experienced decision-making. The results of the Mann-Whitney U test do not show significant differences in the number of errors between novice and expert participants in the quiz (see Table~\ref{h1table}). 

\begin{table}[t!]
\centering
\caption{Mann-Whitney U test results for correctness of decisions.}
\begin{tabular}{@{}lccc@{}}
\toprule
Comparison & U-Statistic & P-Value & Z-Value \\ \midrule
Expert\_15 vs Novice\_15 & 39.5 & 0.433 & -0.794 \\
Expert\_32 vs Novice\_32 & 41.0 & 0.496 & -0.680 \\
Expert\_35 vs Novice\_35 & 50.0 & 1.0 & 0.0 \\ \bottomrule
\end{tabular}
\label{h1table}
\end{table}

\paragraph{Interpretation:} The lack of significant differences in the number of errors suggests that both groups faced similar challenges. Both groups struggled most with Characteristic 35, likely due to the complexity of using two gauges, indicating that this task was equally challenging regardless of expertise.

\subsection{Visual Scanning Time (H2)}
Visual scanning time was measured by tracking the collision of HoloLens 2 gaze rays with scanning markers on various gauges at a sampling rate of 30 Hz (Figure~\ref{visual-gaze}). This data was analyzed to assess the scanning patterns and the decision-making efficiency of the participants. The experts demonstrated a consistent gaze distribution across the gauges, indicating a high level of familiarity. In contrast, novices exhibited higher percentages of gaze on certain gauges, suggesting a learning phase marked by frequent checks and broader, less predictable scanning patterns, indicative of exploratory behavior and uncertainty. The detailed findings and statistical analyzes are as follows (see Table~\ref{tab:ttest_results}): 

\begin{itemize}

\item \textit{Characteristic 15, internal micrometer}: A t-statistic of 2.548 and a \textit{p}-value of 0.020 indicate a significant difference between experts and novices, with experts' gaze time suggesting greater familiarity and clear decision-making.

\item \textit{Characteristic 32, electronic micrometer}: A t-statistic of 3.506 and a \textit{p}-value of 0.0025 indicate a significant difference, with experts showing greater familiarity with the electronic micrometer.

\item \textit{Characteristic 35, electronic caliper and height gauge}: Experts exhibited more evenly distributed scanning time on the electronic caliper and height gauge, with a t-statistic of 2.862 and a \textit{p}-value of 0.010 indicating significance for the caliper. However, for the height gauge, the t-statistic of 1.633 and a \textit{p}-value of 0.120 show no significant difference, likely due to the complexity of using two gauges and varying participant experience.

\end{itemize}

\begin{table}[ht]
\centering
\caption{Independent two-sample t-test for visual scanning time.}
\begin{tabular}{@{}lcc@{}}
\toprule
\textbf{Characteristic \#, Gauge} & \textbf{t-statistic} & \textbf{p-value} \\ 
\midrule
15, internal micrometer & 2.548 & 0.020 \\
32, electronic micrometer & 3.506 & 0.003 \\
35, electronic caliper  & 2.862 & 0.010 \\
35, height  gauge & 1.633 & 0.120 \\
\bottomrule
\end{tabular}
\label{tab:ttest_results}
\end{table}

\paragraph{Interpretation:} 
The analysis identified clear differences in visual scanning and decision-making between experts and novices. Experts demonstrated efficient gaze patterns, indicating familiarity with the tools, which contributed to faster and more accurate decisions. In contrast, novices exhibited broader and less predictable scanning patterns, reflecting their learning stage and associated uncertainty, which resulted in longer decision-making times. The significant differences in scanning time for the internal micrometer (Characteristic 15) and the electronic micrometer (Characteristic 32) highlight the contrast between novices and experts. For Characteristic 35, the more evenly distributed scanning time by experts on the electronic caliper and height gauge suggests greater efficiency. These findings suggest that experts' familiarity with key gauges contributes to their efficient scanning patterns, whereas novices' broader patterns indicate a need for thorough verification and continued learning.

\subsection{Help-Seeking Behavior (H3)}

\paragraph{Session 1: Full AR Instructions}
In Session 1, participants were instructed to begin using the full AR instructions. All novice participants used the full instructions, while 4 expert participants relied on short descriptions instead. The remaining experts used the full instructions similarly to novices. To compare usage patterns and validate the hypothesis that novices and experts differ in their help-seeking behavior, we analyzed the time spent reviewing instructions and the frequency of use of instructional features, including step-by-step 3D animations.

An independent sample t-test showed a significant difference in the time spent reviewing the instructions for Characteristic 5 ($t = -2.781$, $p = 0.012$), with experts spending less time than novices. No significant differences were found for Characteristics 13 ($t = -1.887$, $p = 0.077$) and 6 ($t = -2.104$, $p = 0.050$). The frequency of instruction use showed significant differences across all characteristics: for Characteristic 5 ($t = -2.449$, $p = 0.037$); for Characteristic 13 ($t = -2.449$, $p = 0.037$); and for Characteristic 6 ($t = -2.449$, $p = 0.037$), with novices using the instruction features more frequently than experts.

\paragraph{Session 2: Short AR Instructions}

In Session 2, participants were instructed to use short instructions and only refer to the full instructions if needed. The data revealed varying usage patterns. The t-test for Characteristic 15 did not show significant differences in the time spent reviewing instructions between experts and novices ($t = -0.135$, $p = 0.895$). However, significant differences were found for Characteristics 32 ($t = -2.881$, $p = 0.014$) and 35 ($t = -3.381$, $p = 0.007$), with novices spending more time on these instructions. The frequency of instruction use also showed no significant differences for Characteristic 15 ($t = 0.573$, $p = 0.574$), but significant differences were observed for Characteristics 32 ($t = -2.787$, $p = 0.012$) and 35 ($t = -3.680$, $p = 0.003$), indicating more frequent use by novices.

\paragraph{Interpretation}
The findings from Session 1 suggest that novices relied more heavily on detailed instructions, particularly for Characteristic 5, spending more time and using the instructions more frequently than experts. Although no significant differences were found for Characteristics 13 and 6 in terms of time spent, novices consistently used instructions more frequently. In Session 2, a similar pattern emerged, with significant differences in time spent and frequency of use for Characteristics 32 and 35, where novices again relied more on the instructions. These characteristics involved more specialized gauges and multiple gauges, which may have contributed to novices needing additional guidance. In general, the results indicate that novices are more dependent on help-seeking behavior and detailed instructions, whereas experts are more confident and efficient. This suggests the need for instructional designs that provide structured support for novices while allowing experts to take advantage of their experience.

\subsection{Effect of Previous Experience (H4)}

This section examines how previous experiences influence decision making, tool selection, and adaptation to new tasks for individuals with varying levels of expertise—novices and experts. Descriptive statistics were used to explore the relationships between previous experiences and metrics such as task performance and confidence levels. These metrics offer insights into how each group leverages their experience during task performance. The confidence and preparedness ratings of the post-questionnaire, based on questions such as 'How effectively did you use your previous experiences (e.g., co-ops, internships, projects) to solve problems in the tasks assigned to you' and 'How effective were the tasks in Session 1 at preparing you for the tasks in Session 2?” were rated on a scale from 0 to 10, with 10 being the highest.

\begin{table}[b!]
\centering
\caption{Hypothesis testing results for prior experiences.}
\begin{tabular}{lcc}
\toprule
\textbf{Comparison} & \textbf{t-statistic} & \textbf{p-value} \\
\midrule
Prior Experiences (T-test) & 3.987 & 0.0019 \\
\bottomrule
\end{tabular}
\label{tab:h4_table}
\end{table}

Data analysis centered on the evaluation of H4, suggesting that previous experiences significantly affect decision-making, tool selection, and adaptation to new tasks among novices and experts. The responses of 10 experts and 10 novices were analyzed, detailing their confidence levels in two sessions, the perceived effectiveness of the tasks in session 1 to prepare them for session 2, and their previous experiences. A t-test was performed to test this hypothesis, producing a t-statistic of 3.987 and a p-value of 0.0019, indicating a significant difference in previous experiences between experts and novices (Table~\ref{tab:h4_table}).

\paragraph{Interpretation:} The results show that prior experience significantly impacts decision-making, tool selection, and task adaptation. Experts, with their more extensive and consistent experience, are better equipped to make informed decisions, choose appropriate tools, and quickly adapt to new tasks, while novices tend to rely more on trial and error. The statistical evidence supports the hypothesis that experts handle these aspects more effectively than novices.

\subsection{Stress Management (H5)}

This section explores whether experts experience lower stress levels compared to novices and investigates the effectiveness of stress management in relation to physiological signals such as GSR and NASA-TLX. The goal is to understand whether expertise influences stress responses and decision-making efficiency under pressure.

\paragraph{GSR Metrics:} Analysis of GSR data revealed no statistically significant differences between the expert and novice groups in both decision-making sessions. In Session 1, the Mann-Whitney U test for amplitude resulted in $U = 38.0$ with a $p = 0.385$, and for peaks per minute, the test yielded $U = 40.0$ with a $p = 0.473$. Similarly, in Session 2, the Mann-Whitney U test for amplitude showed $U = 47.0$ with a $p = 0.850$, and for peaks per minute, the test produced $U = 53.5$ with a $p = 0.820$. These $p$ values are all above the conventional significance threshold of 0.05, indicating that the variations observed between experts and novices in both amplitude and peaks per minute are not statistically significant. Consequently, these results suggest that the physiological signals measured by GSR, specifically amplitude and average peak per minute, do not exhibit significant differences between the expert and novice groups in the context of decision-making tasks. This lack of significant variation implies that both groups experienced similar levels of physiological stress, as indicated by their GSR responses.

After analyzing the GSR metrics, including the GSR peaks per minute and the amplitude, we did not find significant differences between experts and novices. Therefore, we decided to further evaluate the NASA-TLX scores to understand the relationship between subjective and objective measures of stress.

\paragraph{NASA-TLX Scores:}
The NASA-TLX scores revealed that both novice and expert groups experience a cognitive load, which remains challenging even for experienced individuals. This finding aligns with the concepts discussed in \cite{wickens2004introduction} regarding workload and expertise in human factors engineering. Experts reported higher physical demands, potentially indicating more intensive participation in the physical aspects of the task. They also experienced higher temporal demands, possibly due to an increased awareness of time constraints or higher performance expectations, as suggested by \cite{endsley1995toward} about situation awareness in dynamic systems.

Experts perceived their performance as significantly higher than that of novices ($t = -2.268$, $p = 0.035$, reflecting their greater skill and efficiency in managing complex tasks. The NASA-TLX performance subscale assesses how successful individuals believed they were in accomplishing the goals of the task set by the experimenters or themselves and how satisfied they were with their performance in accomplishing these goals. This phenomenon is discussed in \cite{ericsson1996expert} in their review of expert performance. Despite facing higher mental and physical demands, experts reported exerting less effort, suggesting that experience can make tasks less strenuous through more efficient strategies and greater familiarity with the task. In addition, experts experienced lower levels of frustration, indicating that experience helps to effectively manage stress and challenges. This supports the notion that expertise facilitates adaptation to task constraints. This trend toward reducing frustration among experts is evidenced by the near-significant p-value ($t = -2.001$, $p = 0.066$).

\begin{figure}[h!]
    \center
    \includegraphics[width=1\linewidth]{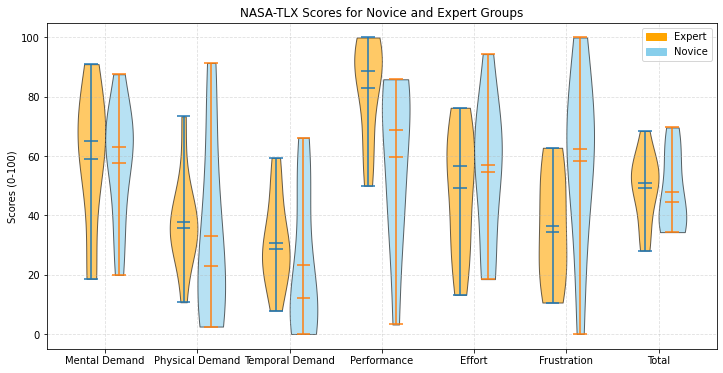}
    \caption{Novice and expert NASA-TLX scores.}
    \label{fig:nasa-tlx}
\end{figure}

\paragraph{Correlation Between NASA-TLX Scores and GSR Metrics:}
Understanding the relationship between subjective and objective measures of stress is crucial. Subjective measures such as NASA-TLX scores for performance and frustration provide insight into individuals' perceived workload and emotional responses. Objective measures like GSR metrics (amplitude and peaks per minute) offer physiological data related to stress and arousal levels. By correlating these measures, we can determine how subjective experiences of stress and workload relate to physiological responses, revealing if individuals' perceptions align with their physiological states.

Pearson's correlation analysis was performed between NASA-TLX scores (Performance and Frustration) and GSR metrics for both experts and novices, due to the significant results from the t-test for Performance and the near-significant result for Frustration in NASA-TLX. The results are summarized in Figure~\ref{fig:nasa-tlx}. For experts, the correlations were as follows: Performance and GSR Amplitude ($r = 0.347$, $p = 0.326$), Performance and GSR Peaks per Minute ($r = 0.161$, $p = 0.656$), Frustration and GSR Amplitude ($r = 0.026$, $p = 0.943$), and Frustration and GSR Peaks per Minute ($r = 0.099$, $p = 0.786$). For novices, the correlations were as follows: Performance and GSR Amplitude ($r = 0.723$, $p = 0.018$), Performance and GSR Peaks per Minute ($r = 0.525$, $p = 0.120$), Frustration and GSR Amplitude ($r = -0.290$, $p = 0.417$), and Frustration and GSR Peaks per Minute ($r = -0.115$, $p = 0.752$).

\begin{figure}[b!]
    \center
    \includegraphics[width=1\linewidth]{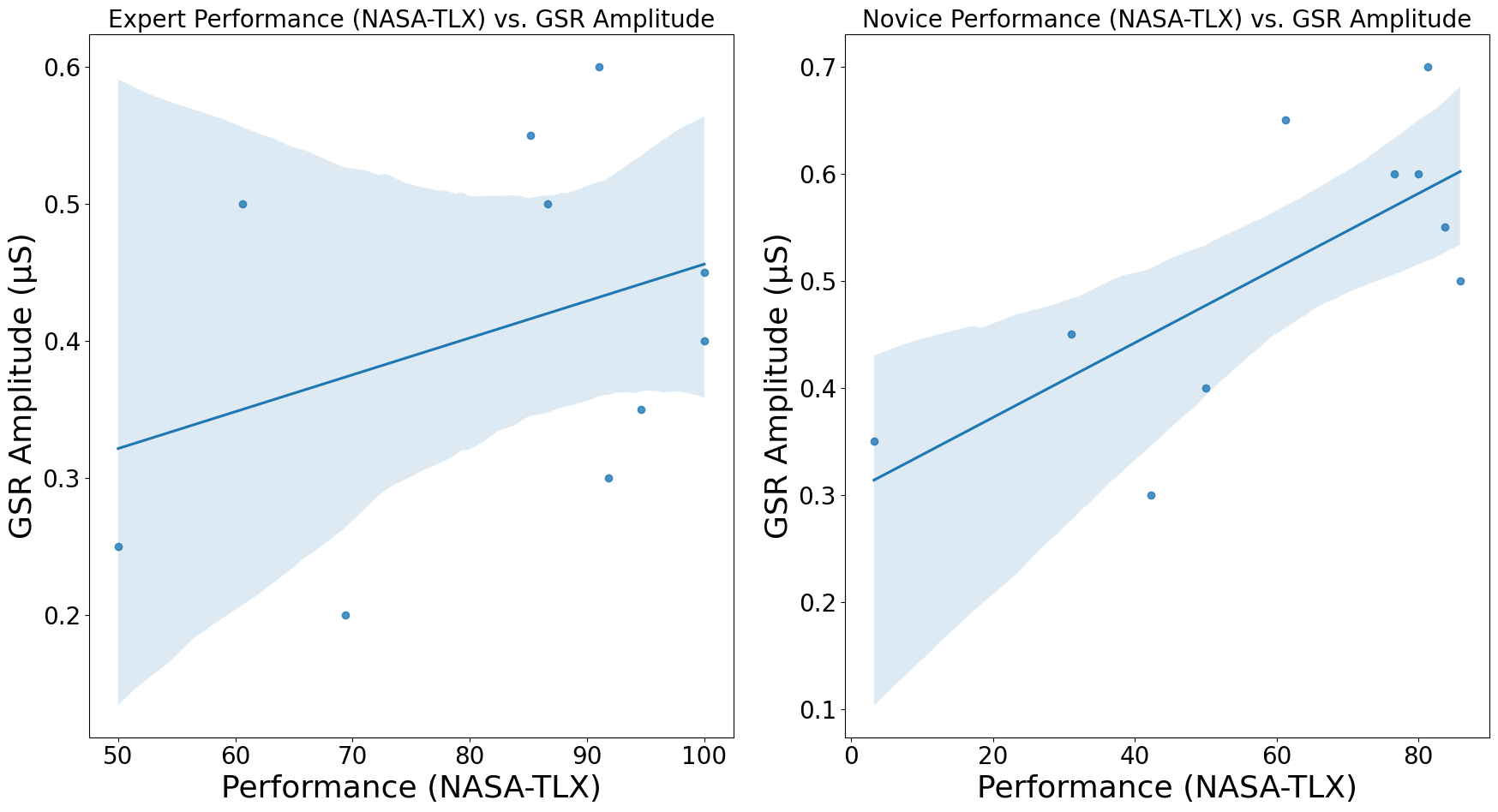}
    \caption{Scatter plots with regression lines showing the relationship between performance and GSR amplitude for experts (left) and novices (right).}
    \label{fig:H5}
\end{figure}

The novices showed a positive correlation between perceived performance and GSR amplitude ($r = 0.723$, $p = 0.018$), indicating that higher perceived performance is associated with increased physiological stress (Figure~\ref{fig:H5}). Experts showed weak correlations, suggesting a more effective stress modulation. This analysis indicates the need for more research to determine why experts are better at mitigating stress during tasks. 

\paragraph{Interpretation:} The analysis revealed that both experts and novices experienced similar mental demand during decision-making tasks. However, experts reported higher physical and temporal demands, perceived their performance as higher, and reported lower effort and frustration compared to novices. The correlation analysis showed that experts had negative correlations between GSR amplitude and several NASA-TLX subscales (e.g., physical demand, temporal demand), indicating that higher perceived stress was associated with lower physiological responses. This suggests that experts can better modulate their physiological stress. In contrast, novices exhibited positive correlations, indicating that higher perceived stress was associated with higher physiological responses, reflecting greater physiological stress.

The results suggest that stress management improves with experience, enabling experts to better control their physiological responses under stress. In contrast, novices tend to encounter higher physiological stress in comparable situations. The potential factors for the higher temporal demand reported by experts could be their relative higher expectations from themselves. The greater physical demand might be attributed to their familiarity with performing the task without the additional setup of the experiment (e.g., headset, GSR equipment), which could introduce physical discomfort or distraction. Further experiments and analyses are indeed needed to measure these potential factors more accurately and explore additional variables that can influence stress management and physiological responses in decision-making tasks.

The study used a correlation analysis GSR and NASA-TLX scores to assess the relationship between physiological stress and subjective workload perception. The validity of this methodological approach is supported by prior research, which has demonstrated the efficacy of similar analyses in the context of workload assessment. For example, Delliaux \textit{et al.} highlight that GSR, as a measure of autonomic nervous system activity, can be effectively correlated with subjective workload assessments such as NASA-TLX to gauge cognitive and emotional stress levels during task performance \cite{delliaux2019mental} Furthermore, the broader literature on workload measurement frequently employs physiological indicators such as heart rate variability (HRV) and GSR in conjunction with subjective tools like NASA-TLX. This dual approach is validated by studies showing consistent correlations between these measures in various task environments.

\subsection{Accuracy of Inspection (H6)}

The total correct inspections for both Session 1 and Session 2 (accept or reject) show a significant difference between novices and experts. The results of the t-test are $t = 3.07$, $p = 0.007$, which is below the significance threshold of 0.05. This indicates that experts generally perform better on parts inspection compared to novices. Although individual tasks do not show significant differences in most cases, the overall performance metric reveals that experts perform better than novices in inspection tasks. This suggests that expertise plays a role in the overall accuracy of the inspections of the parts. Accuracy is measured by analyzing task performance and report scores that reflect the accuracy of task execution and inspection results. An error is defined as any deviation from the correct inspection procedure or an incorrect inspection result. Thus, higher accuracy corresponds to fewer errors and more correct inspections.

\paragraph{Interpretation:} There is a significant difference in inspection accuracy between novices and experts, and experts perform better overall. Although individual tasks did not show significant differences, the aggregate performance metric highlights the importance of expertise in inspection tasks. This suggests that experience and skill substantially enhance overall inspection accuracy.

\subsection{Dexterity (H7)}

\begin{table}[t!]
    \centering
    \caption{Pause metrics comparison between novices and experts.}
    \label{tab:hesitation_metrics}
    \begin{tabular}{lcc}
        \toprule
        \textbf{Metric} & \textbf{Novices} & \textbf{Experts} \\
        \midrule
        Frequency & 41,002 & 78,261 \\
        Ratio & 0.75 & 0.67 \\
        Mean Duration (seconds) & 16.2 & 12.7 \\
        Median Duration (seconds) & 4 & 2 \\
        Longest Duration (seconds) & 216 & 287 \\
        \midrule
        \multicolumn{3}{c}{\textbf{Duration Distribution Overview}} \\
        \midrule
        0-10 seconds (\%) & 70.82 & 81.08 \\
        11-30 seconds (\%) & 17.52 & 13.33 \\
        31-60 seconds (\%) & 7.27 & 3.52 \\
        Over 60 seconds (\%) & 4.39 & 2.07 \\
        \bottomrule
    \end{tabular}
\end{table}

We observe that experts have a higher frequency of pauses (78,261) compared to novices (41,002), as shown in Table~\ref{tab:hesitation_metrics}. This suggests that experts engage in more frequent but shorter reflective pauses to optimize their performance. 

\begin{itemize}
[itemsep=-0.25ex]
    \item \textit{Frequency comparison}: The results of the chi-square test showed a significant difference in the frequency of pauses between novices and experts ($\chi^2 = 15126.53$, $p < 0.001$).

    \item \textit{Ratio comparison}: The t-test for the ratio of pause time to task time between novices and experts indicated a significant difference ($t = 336.21$, $p < 0.001$), with experts having a lower ratio (0.67) compared to novices (0.75), suggesting more efficiency in their task execution.
    
    \item \textit{Mean and median duration comparison}: The t-test for mean duration showed a significant difference ($t = 182.92$, $p < 0.001$), with experts having shorter mean pause durations (12.7 seconds) compared to novices (16.2 seconds). The Mann-Whitney U test for median duration also indicated a significant difference ($U = 820712.5$, $p < 0.001$), with experts having a median duration of 2 seconds compared to 4 seconds for novices.
    
    \item \textit{Duration distribution comparison:} The chi-square test for the distribution of pause durations across different categories (0-10 seconds, 11-30 seconds, 31-60 seconds, over 60 seconds) indicated significant differences between novices and experts ($p = 0.429$).
    
    \item \textit{Longest pause duration comparison}: The t-test for the longest pause duration showed a significant difference ($t = 94.59$, $p < 0.001$), with experts having a longer maximum pause duration (287 seconds) compared to novices (216 seconds).

\end{itemize}

\paragraph{Interpretation:} The analysis of pause metrics suggests that dexterity, as evidenced by pauses during manipulation tasks, varies significantly between novices and experts. Experts not only pause more frequently, but also manage to keep their pauses shorter on average, indicating a higher level of proficiency and efficiency in handling tasks. The lower ratio of pause time to task time in experts (0.67) compared to novices (0.75) further emphasizes this efficiency. These findings align with theories of expertise that propose that expert performance is characterized by the ability to quickly navigate or recover from potential errors or uncertainties through more effective cognitive and physical adjustments. These insights into hand hesitation and pauses can help to understand the behavioral nuances that distinguish expert performers from novices, potentially guiding training methods to reduce pause durations and frequency in novice learners. For this particular case, we have examined only the wrist joint. We would like to expand to more joints and explore more fine-grained action movements to see how novice and expert groups' hand movements differ in future studies. Significant differences in hand movement patterns were observed between novices and experts in psychomotor tasks. The experts exhibited shorter and more frequent pauses, indicating quicker decision making and corrections.
\subsection{Task Execution Efficiency (H8)}

\begin{figure}[h!]
    \center
    \includegraphics[width=1\linewidth]{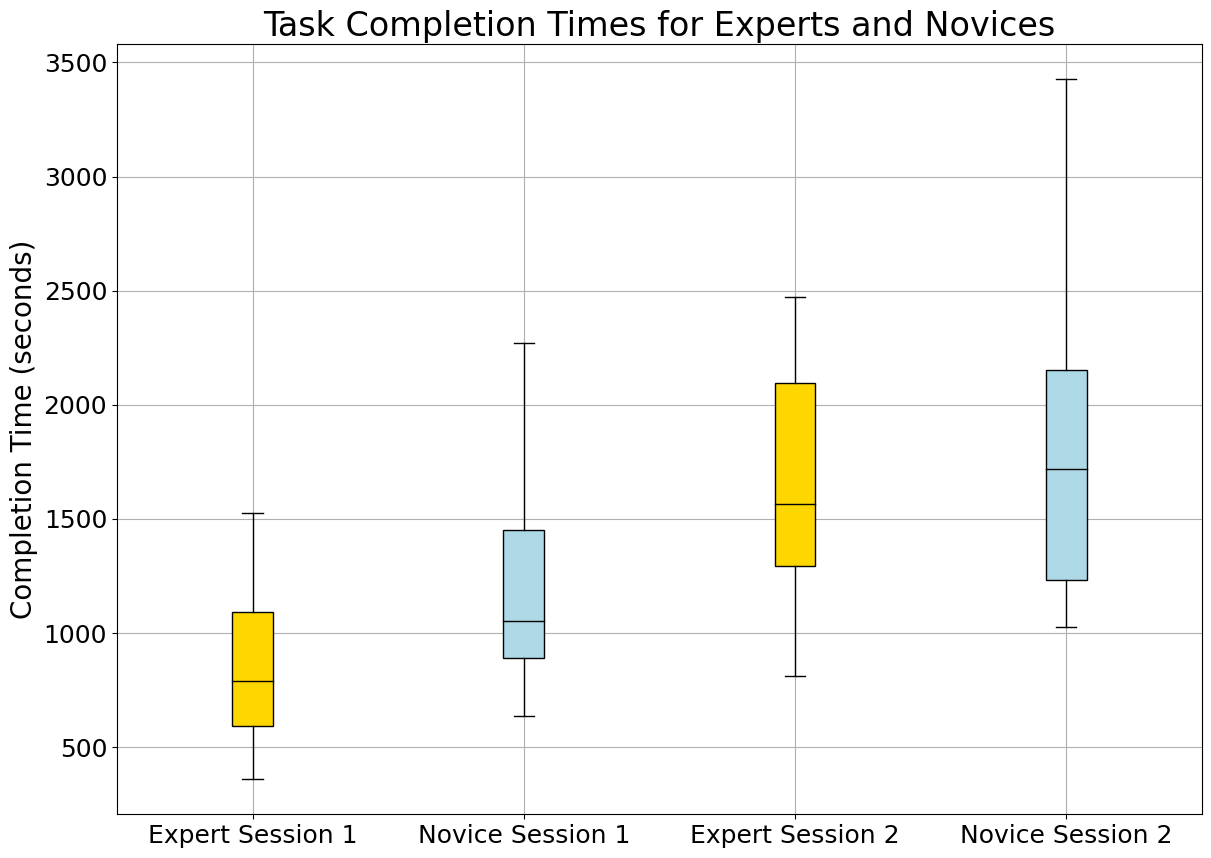}
    \caption{Differences between expert and novice task completion times in Sessions 1 (statistically significant) and 2 (not significant).}
    \label{fig:completion-time}
\end{figure}

The objective of this study was to compare the efficiency of task execution between experts and novices over two sessions, measured by task completion time. t tests were performed to compare the groups. In Session 1, the t-test produced a t statistic of 2.768 (p = 0.013), indicating a significant difference, with experts completing tasks faster. In Session 2, the t-test produced a t statistic of 0.676 (p = 0.507), indicating that there were no significant differences. Although experts generally completed tasks faster than novices, significant differences were found only in Session 1. The variability in task completion times, particularly in Session 2, suggests that factors other than skill level may influence performance.

\paragraph{Interpretation:} Experts completed tasks faster than novices, but significant differences were only found in Session 1 (Figure~\ref{fig:completion-time}). This supports previous findings that expertise improves task performance. In Session 2, no significant differences were found, suggesting that other factors, such as task complexity or learning effects, may impact time completion. The results show that while expertise enhances efficiency, it is influenced by multiple factors beyond skill level.

\section{Conclusions}
\label{conclusion}

This study aimed to explore the distinctions between novice and expert performance in AR-guided industrial psychomotor tasks, focusing on two primary dimensions of expertise: decision-making and technical proficiency. By examining the results across multiple hypotheses, we identified significant differences and commonalities that enhance our understanding of expertise in industrial settings. The findings of this study highlight the multifaceted nature of expertise in industrial psychomotor tasks. Experts not only demonstrate higher technical proficiency, but also exhibit more efficient decision-making processes. However, many hypothesized differences, such as independence from instructions or mental demand, proved to be insignificant. These insights can inform the design of adaptive AR systems that dynamically tailor instructions and interventions to the user's level of expertise, providing personalized guidance and support to enhance training and performance.  

The psychomotor task utilized in this study represents only one of many possible tasks. The results obtained may vary depending on the nature and complexity of different psychomotor tasks, as these factors could significantly influence both physiological and subjective responses. Our findings should be viewed in the context of this specific task, and further research is needed to explore whether similar patterns emerge across a broader range of psychomotor tasks. Future research should focus on larger sample sizes and diverse tasks to further validate these findings. In addition, exploring cognitive and motivational influences on performance can provide a more comprehensive understanding of expertise. Using the multimodal data captured from AR devices and wearables, future studies can refine expertise estimation models, ultimately contributing to the development of adaptive AR systems that optimize industrial training and assistance. In conclusion, this research lays the foundation for improving AR technologies in industrial settings by elucidating the critical dimensions of expertise and their impact on task performance. The continued exploration of these dimensions will be essential for advancing personalized and effective training solutions in industrial psychomotor tasks.

\acknowledgments{
This material is based on work supported by the National Science Foundation grants FW-HTF-R-2128743 and RITEL-2302838. Any opinions, findings, or conclusions expressed in this material are those of the authors and do not reflect the views of the National Science Foundation.}

\bibliographystyle{unsrt}

\bibliography{template}
\end{document}